\documentclass[conference]{IEEEtran}
\IEEEoverridecommandlockouts
\usepackage{cite}
\usepackage{verbatim}

\ifdefined\autocite
\else
  \newcommand{\autocite}{\cite}
\fi

\usepackage{amsmath,amssymb,amsfonts}
\usepackage{algorithmic}
\usepackage{graphicx}
\usepackage{siunitx}
\usepackage{textcomp}
\usepackage{xcolor}
\usepackage{hyperref}
\usepackage[nolist]{acronym}
\usepackage{color,soul}

\def\BibTeX{{\rm B\kern-.05em{\sc i\kern-.025em b}\kern-.08em
    T\kern-.1667em\lower.7ex\hbox{E}\kern-.125emX}}
\begin{document}

\title{PEtra: A Flexible and Open-Source PE Loop Tracer for Polymer Thin-Film Transducers\\
\thanks{The authors acknowledge support from the ETH Research Grant ETH-C-01-21-2 (Project ListenToLight). \\
Corresponding author: {christoph.leitner@iis.ee.ethz.ch}}}

\author{
\IEEEauthorblockN{
    Marc-Andre~Wessner\IEEEauthorrefmark{2}\IEEEauthorrefmark{1},
    Federico~Villani\IEEEauthorrefmark{2}\IEEEauthorrefmark{1}, 
    Sofia~Papa\IEEEauthorrefmark{3}, 
    Kirill~Keller\IEEEauthorrefmark{4},
    Laura~Ferrari\IEEEauthorrefmark{3},\\ 
    Francesco~Greco\IEEEauthorrefmark{3}\IEEEauthorrefmark{4}, 
    Luca~Benini\IEEEauthorrefmark{2}\IEEEauthorrefmark{5},
    Christoph~Leitner\IEEEauthorrefmark{2}}

\IEEEauthorblockA{
    \IEEEauthorrefmark{1} Authors contributed equally to this work. \\
    \IEEEauthorrefmark{2}Department of Information Technology and Electrical Engineering, ETH Zurich, Switzerland\\
    \IEEEauthorrefmark{3}The BioRobotics Institute, School of Advanced Studies, Pisa, Italy \\
    \IEEEauthorrefmark{4}Department of Physics, Graz University of Technology, Graz, Austria \\
    \IEEEauthorrefmark{5}DEI, University of Bologna, Bologna, Italy\\}
}

\maketitle

\begin{acronym}
    \acro{ldo}[LDO]{low-dropout regulator}
    \acro{dip}[DIP]{dual in-line package}
    \acro{Z}[Z]{impedance}
    \acro{L}[L]{inductive}
    \acro{C}[C]{capacitive}
    \acro{R}[R]{resistive}
    \acro{tia}[TIA]{transimpedance amplifier}
    \acro{snr}[SNR]{signal-to-noise ratio}
    \acro{pvdf}[P(VDF-TrFE)]{Poly(vinylidene fluoride-tetrafluoroethylene)}
    \acro{us}[US]{ultrasound}
    \acro{pzt}[PZT]{lead zirconate titanate}
    \acro{pe}[PE]{polarization-electric field}
    \acro{DIP}{dual in-line package}
    \acro{SNR}{Signal to Noise Ratio}
    \acro{pcb}[pcb]{printed circuit board}
\end{acronym}


\begin{abstract}
Accurate characterization of ferroelectric properties in polymer piezoelectrics is critical for optimizing the performance of flexible and wearable \ac{us} transducers, such as screen-printed \ac{pvdf} devices. Standard charge measurement techniques, like the Sawyer–Tower circuit, often fall short when applied to ferroelectric polymers due to low-frequency leakage. In this work we present \textit{PEtra}, an open-source and versatile \ac{pe} loop tracer. \textit{PEtra} employs a \ac{tia} (LMP7721, TI) to convert picoampere-level currents into well measurable voltages, covering a frequency range of \qty{0.1}{Hz} to \qty{5}{Hz} for a gain setting of  $10^7\:V/A$, and \qty{0.1}{Hz} to \qty{200}{Hz} for gain settings between $10^3\:V/A$ to $10^6\:V/A$ (10-fold increments). We demonstrate through simulations and experimental validations that \textit{PEtra} achieves a sensitivity down to \qty{2}{\pico\ampere}, effectively addressing the limitations of traditional charge measurement methods. Compared to the Sawyer-Tower circuit, \textit{PEtra} directly amplifies currents without the need for a reference capacitor. As a result, it is less susceptible to leakage and can operate at lower frequencies, improving measurement accuracy and reliability. \textit{PEtra}'s design is fully open source, offering researchers and engineers a versatile tool to drive advancements in flexible \ac{pvdf} transducer technology.

\end{abstract}


\vspace{0.5pt}
\section{Introduction}
\label{sec:introduction}

Recent advances in medical \ac{us} transducers have led to a range of designs that can bend and conform to body shapes, primarily utilizing \ac{pzt} \mbox{1-3} composite structures embedded in flexible, skin-adhesive matrices~\mbox{\cite{lin2024, wang2022}}. \ac{pvdf}, a lead-free piezoelectric polymer, has emerged as a promising alternative to \ac{pzt} for flexible and wearable \ac{us} transducer designs \cite{van_neer_flexible_2024, keller_fully_2023}. While the transmit performance of \ac{pvdf} is lower than that of piezoelectric ceramics, its inherent flexibility and bendability make it ideal for wearable applications. Several factors make the usage of \ac{pvdf} advantageous for fast iterative research. Firstly, the acoustic impedance of \ac{pvdf} closely matches that of human tissue \cite{rathod_review_2020}, effectively eliminating the need for additional matching layers. Moreover, the compatibility of \ac{pvdf} with solution-based deposition methods allows it to be used in printing and additive manufacturing techniques \cite{wagle_ultrasonic_2013}. Screen-printing of \ac{pvdf} is particularly promising due to its simple, cost-effective, and scalable production. 
Despite these advantages, screen-printing methods for medical \ac{pvdf} transducers have not yet been thoroughly researched, necessitating significant flexibility to optimize both fabrication processes and transducer architectures \cite{keller_fully_2023}. Achieving this flexibility in design demands equally adaptable and open tools for accurate characterization \cite{leitner_design_2022}. 

Reliable measurement of ferroelectric properties is essential to ensure consistent printing, polarization, and performance throughout the fabrication process. \ac{pe} loops, in particular, are crucial for understanding how polarization in piezoelectrics responds to external electric fields, revealing key material properties like energy storage, domain switching, and mechanical actuation \cite{stewart_ferroelectric_nodate}. The shape and characteristics of these curves - such as remnant polarization, coercive field, saturation polarization, and hysteresis width - are directly linked to the material's electromechanical coupling behavior. However, creating and measuring these fields for \ac{pe} loop tracing represents a substantial technical challenge in ferroelectric polymers~\cite{wegener_polarization-electric_2008}. These materials necessitate higher electric fields and lower characterisation frequencies than piezoelectric ceramics~\cite{qiu_direct_2013}, which is attributed to their elevated coercive fields and prolonged switching times.
In particular, the characterization of these phenomena at low injection frequencies necessitates the ability to precisely measure the quasistatic changes in the piezoelectric element. The well-known Sawyer-Tower circuit fails to analyze these conditions due to its low sensitivity and high leakage \cite{perera_analysis_2021}. While charge measurement devices are well described \cite{stewart_ferroelectric_nodate}, open-source equipment for characterizing \ac{pvdf} remains unavailable.

In this context, this paper introduces \textit{PEtra}, an open-source, flexible \ac{pe} loop tracer for characterizing ferroelectric properties of screen-printed \ac{pvdf} films. Our device utilizes a current measurement principle and offers wide frequency adaptability. The circuit amplifies small currents (in the low \qty{}{\pico\ampere} range) into well measurable voltages (over \qty{1}{\milli\volt}) with low noise (\ac{snr} of \qty{6}{\decibel}) for input frequencies from \qty{0.1}{\hertz} (quasistatic) up to \qty{200}{\hertz}. This paper outlines the design, implementation, and validation of our \ac{pe} loop tracer. We provide a versatile open-source tool for researchers and engineers working on \ac{pvdf} with full public access to the source data available at: \mbox{\url{https://github.com/pulp-bio/PEtracer}}.

\begin{figure}[bt]
    \centering
      \includegraphics[width=\columnwidth, clip, trim={0 0 0 0}]{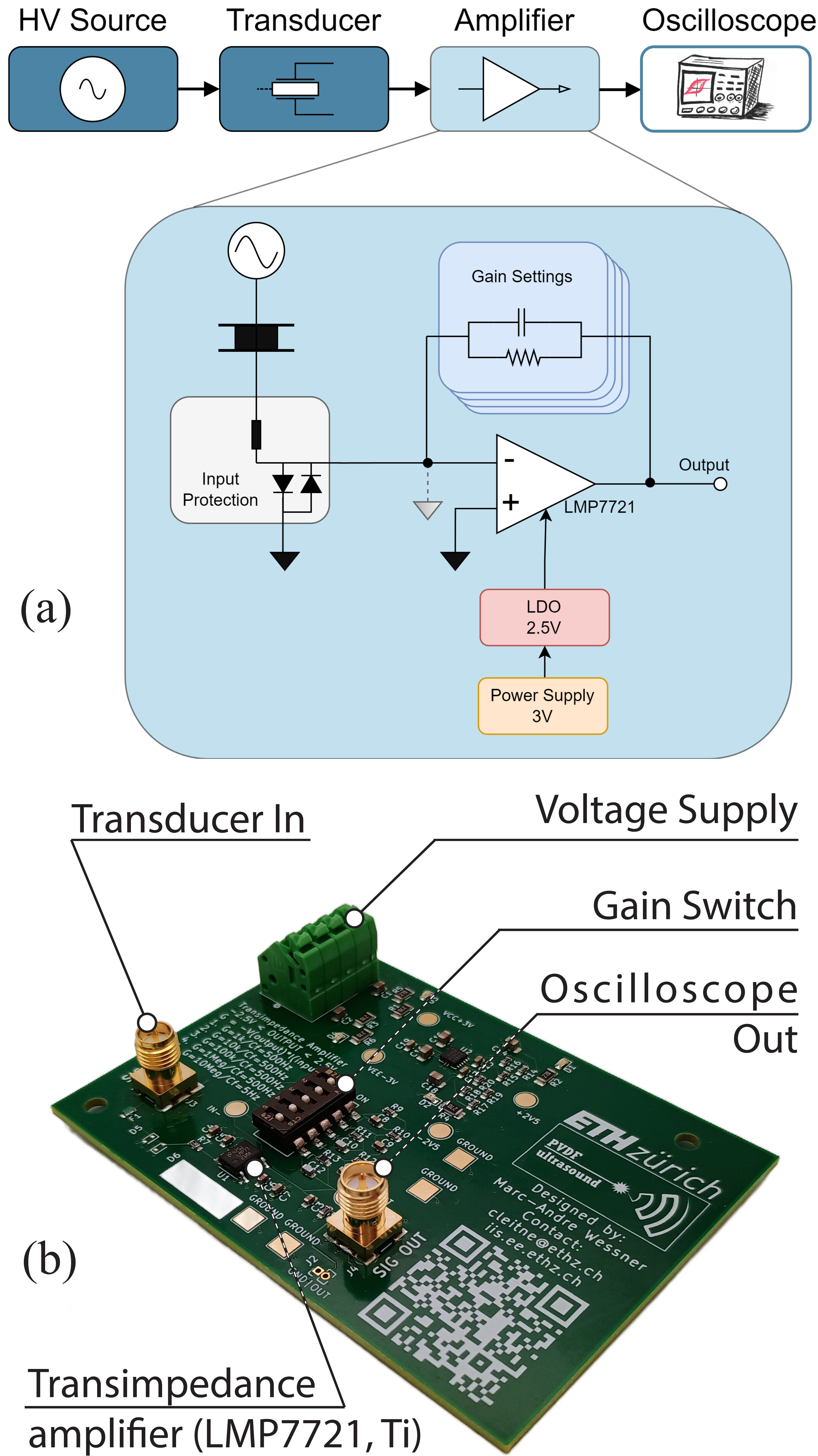}
    \caption{(a) shows a block diagram of the \ac{pe} loop tracer in a transducer testing setup. A high-voltage source excites the transducer, generating a current amplified and measured by an oscilloscope. The light blue square details \textit{PEtra}'s input protection circuit (fuse and clipping diodes), selectable feedback network (resistors and low-pass capacitors), and the LDO connecting the TIA to the power supply. (b) shows the physical printed circuit board of \textit{PEtra} with five marked interaction points. "Transducer In" indicates the current input and virtual ground for transducer characterization, while "Voltage Supply" powers the device. "Gain Switch" refers to the DIP switch for selecting gain, and "Oscilloscope Out" provides the signal output. The LMP7721 TIA, the core of \textit{PEtra}, is also highlighted.}
    \label{fig:circuit}
\end{figure}

\section{Materials and Methods}
\label{sec:methods}
In order to address the low characterisation frequency demands for ferroelectric polymers, we selected a \ac{tia} current measurement method. This section introduces our proposed circuit and details the optimisation of the \ac{tia} for capturing \ac{pe} loops from ferroelectric polymers over a wide range of transducer designs.

\subsection{Circuit Design }
The core of our measurement circuit (Figure \ref{fig:circuit}) is a \ac{tia} based on the LMP7721 operational amplifier (Texas Instruments, USA), chosen for its ultra-low input bias current of \qty{3}{fA}. A \ac{ldo} provides the op-amp with a stable dual power supply of  \(\pm2.5V\), minimizing power supply fluctuations and ensuring optimal performance.
To safeguard the op-amp from damage caused by high polarization conditions (e.g., transducer failure and short circuits) we added a \qty{10}{\milli\ampere} fuse to the input path. Adjustable gain settings are implemented using a \ac{DIP} switch that selects among various feedback resistors ($R_f$) and capacitors ($C_f$). This configuration allows for transimpedance gains ranging from $10^3$ to $10^7 V/A$ in decade steps. The system is designed to amplify currents within a frequency range of \qty{0.1}{Hz} (quasistatic) to \qty{200}{Hz} for decade steps 1 to 4 and \qty{0.1}{Hz} to \qty{5}{Hz} for the highest gain setting.

\subsection{Transducer Model}
For the system parameterization, we incorporated a screen-printed \ac{pvdf} transducer on a polyimide substrate, as detailed in \cite{keller_fully_2023}. We modelled the transducer using the Butterworth-Van Dyke equivalent resonator circuit model \cite{van_dyke_piezo-electric_1928}, simplified to a series RC circuit in order to align with the requirements of our low-frequency application. At frequencies below \qty{1}{\kilo\hertz}, the model's inductive components can be considered negligible, with resistive and capacitive elements primarily influencing the impedance. 

To accurately determine the RC components for our simulation, we measured the resistance and reactance of the \ac{pvdf}) transducer using an LCR meter (E4980AL, Keysight, USA) over a frequency range of $f \in [\qty{20}{Hz}, \qty{1000}{Hz}]$, with 201 linearly spaced measurement points. The measured impedance was then fitted to the RC model, described by the equation $Z = R + 1/j 2 \pi f C$, yielding a resistance of $R=\qty{131.8}{k\Omega} \pm\qty{11.6}{k\Omega}$ and a capacitance of $C=\qty{0.707}{nF} \pm\qty{1.52}{pF}$.

\subsection{Circuit Simulation and Optimization}
To optimize the amplifier's gain settings and feedback loop, we conducted comprehensive simulations to ensure the device's stability and linearity across all measurement conditions. We utilized LTSPICE (Analog Devices, USA), using the manufacturer's SPICE model for the LMP7721 op-amp \cite{TI_lmp7721_2024}, incorporating the RC model for the transducer as described in the previous subsection.

To account for variability in transducer characteristics, we swept the transducer's impedance by a factor of ten above and below the nominal values. Simultaneously, we varied the impedance values of $R_f$ and $C_f$ in 10-fold increments from \qty{1}{\kilo\ohm} to \qty{10}{\mega\ohm}. The feedback capacitor values were selected to achieve a cutoff frequency of \qty{500}{\hertz} for resistances from \qty{1}{\kilo\ohm} to \qty{10}{\mega\ohm} (i.e., gain settings 1-4), and a cutoff frequency of \qty{5}{\hertz} for \qty{10}{\mega\ohm} (i.e., highest gain setting), ensuring optimal bandwidth for the different gain settings. The feedback capacitor value \(C_f\) was obtained for each cutoff frequency by the equation $C_f = 1/2 \pi f_\text{cutoff} R_f$. 

Based on the simulation results, we characterized and optimized each gain step, designing the circuit to measure over a total of nine decades of current magnitudes, ranging from \qty{2}{\pico\ampere} up to \qty{2}{\milli\ampere}, across all gain settings. For each individual gain setting, the circuit can measure over five orders of magnitude of current, allowing for the characterization of a single transducer without change of gain setting.

\subsection{Circuit Characterization}
To assess the accuracy of our device, we simulated the transducer current using a function generator (33600A, Keysight Technologies, USA) connected in series with a \qty{1}{\mega\ohm} resistor and our measurement device. Both the input and output voltages were measured using a digital oscilloscope (InfiniiVision MSOX3024T, Keysight Technologies, USA), and the current flowing into the amplifier was calculated using Ohm's law. To characterize the system, we varied the voltage amplitude of the waveform generator and the gain settings, accumulating 37 measurement points. Direct measurements were conducted only at frequencies of \qty{3}{\hertz} for the highest gain setting and \qty{100}{\hertz} for the lowest gain setting. All other data points (i.e. for the given frequency range of \qty{0.1}{\hertz} to \qty{200}{\hertz}) were derived from simulations. Moreover, due to limitations in the measurement and stimulation devices (i.e. the waveform generator's minimum and maximum voltage amplitudes of \qty{1}{\milli\volt} and \qty{10}{\volt}, respectively), the experimental characterization did not encompass the full sensitivity range of the amplifier. Thus, we used exponential regression to estimate device performance in the upper and lower current regions.

To evaluate the accuracy of our device we analyzed the recorded signals in MATLAB (Mathworks, USA). We first applied a \qty{50}{\hertz} notch filter to the acquired output voltage signals to remove the fundamental line frequency noise in Switzerland. We calculated the \ac{snr} using MATLAB's built-in \textit{snr} function \cite{mathworks_inc_matlab_2024}, which performs a periodogram with a Kaiser window, identifies the fundamental frequency as the largest spectral component, and estimates noise from the median power in noise-only regions. We fitted the measured \ac{snr} values to an exponential regression model $f(x)=ae^{xb}$ to estimate the performance in the upper and lower current regions.

The device's current range limits were set for each gain setting, with the lower bound defined as the current corresponding to an \ac{snr} of \qty{6}{\decibel}, and the upper bound constrained by the maximum output voltage of the \ac{tia}, \qty{2.41}{\volt}.

\begin{figure}[t!]
    \centering
    \vspace{0.2cm}
      \includegraphics[width=\columnwidth, clip, trim={0 0 0 0}]{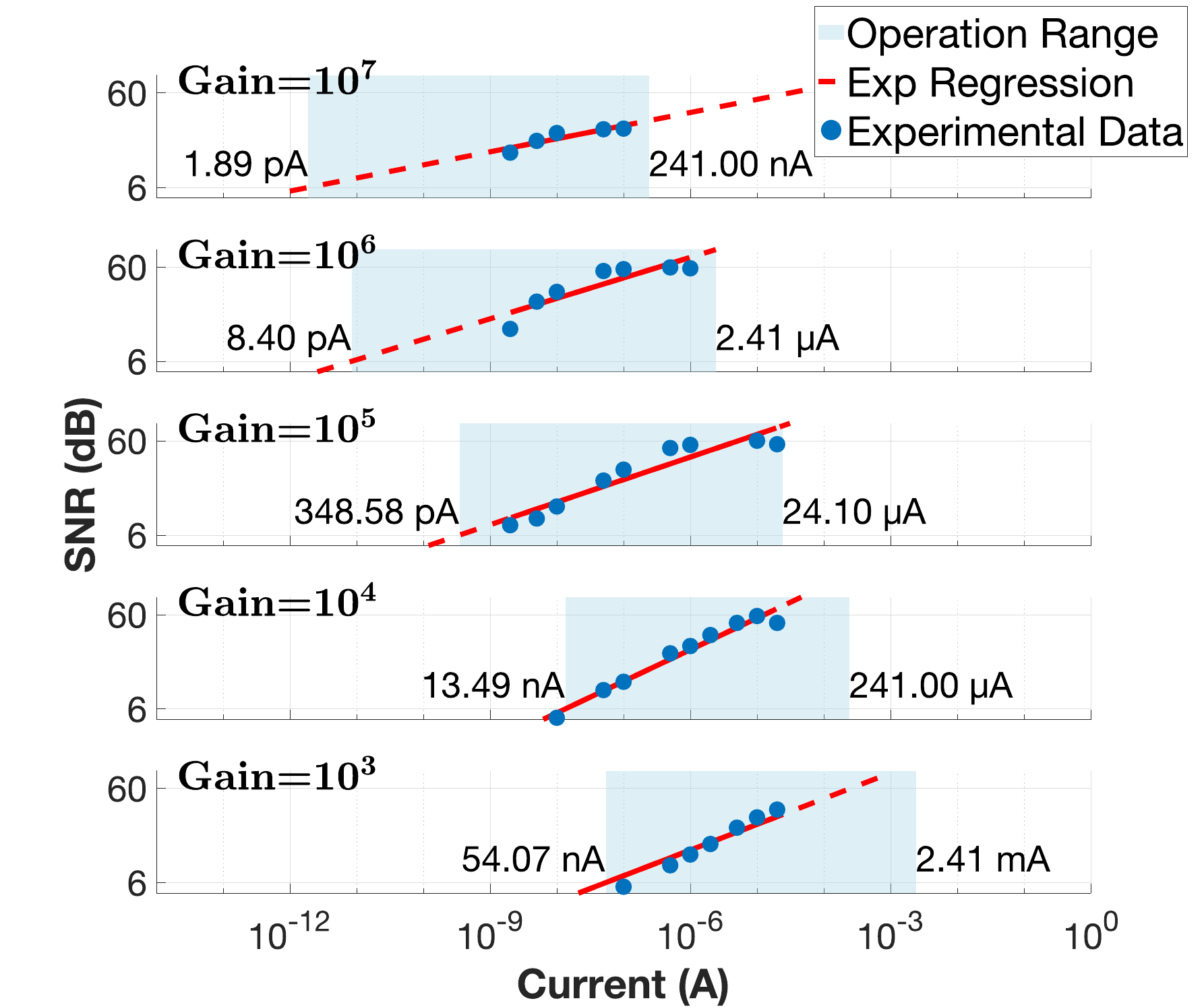}    
      \caption{
      Device characterization results with the input current on the x-axis (logarithmic scale) and the SNR in decibels on the y-axis for each gain setting. Blue dots represent individual current measurements, while the red curve illustrates an exponential regression model extrapolating these currents into lower SNR regions.}
      \vspace{0.2cm}
    \label{fig:accuracy}
\end{figure}
\newpage
\section{Results and Discussion}
\label{sec:results}

Figure \ref{fig:accuracy} illustrates the accuracy limits of our device across the five selectable gain settings. The x-axis represents the input current on a logarithmic scale, while the y-axis shows the \ac{snr} in \qty{}{\decibel} for each gain setting. Blue dots represent individual current measurements, while the red curve depicts an exponential regression model fit extrapolating these currents into lower \ac{snr} regions, illustrating the expected accuracy of our device. Measurements in these regions are not possible in our setup due to the limitations of the signal generator. The operational ranges of the individual settings are highlighted in light blue, with the respective upper and lower current limits indicated at the left and right edges of each highlighted area.

At the highest gain setting, our experimental evaluation demonstrates a sensitivity of \qty{2}{\pico\ampere} with a \ac{snr} of \qty{6}{dB}, enabling characterization at quasistatic states. This capability is crucial for \ac{pvdf} transducers, necessitating low-frequency characterization due to their slow domain wall movement and long switching times. 

Unlike traditional charge amplification methods, our device avoids draining and roll-off effects at low frequencies because it directly amplifies the current without relying on a reference capacitor. The integration of the current signal is performed downstream via software, allowing for preliminary filtering of the raw signal to mitigate noise sources such as mains frequency interference. 
\newpage
\section{Conclusion}
\label{sec:conclusion}

We have developed an open-source, flexible \ac{pe} loop tracer specifically tailored to accurately characterize piezoelectric properties of screen-printed \ac{pvdf} transducers. By employing a \ac{tia} with adjustable gain settings ranging from $10^3 \:V/A$ to $10^7\:V/A$, our device amplifies currents starting from as low as \qty{2}{\pico\ampere}, into easily measurable voltages of \qty{1}{\milli\volt} and upwards with a \ac{snr} exceeding \qty{6}{\decibel}. The amplification spans a broad frequency range: At the highest gain setting and frequencies from \qty{0.1}{\hertz} to \qty{5}{\hertz} our \ac{pe} loop tracer enables accurate characterization under quasistatic conditions inherent to ferroelectric polymers. Additionally, for lower gains and currents above \qty{8.4}{\pico\ampere}, the measurable frequency band extends up to \qty{200}{\hertz}.

Our method circumvents the limitations of common charge measurement techniques by directly amplifying the current without reliance on reference capacitors, thus avoiding issues like signal draining and roll-off at low frequencies. The downstream integration and filtering of the current signal in software enhance measurement accuracy and reliability. 
This work offers an open-source tool that aids researchers and engineers working with ferroelectric polymers in optimizing printing processes, polarization methods, and performance assessments of \ac{pvdf} transducers, thereby encouraging further development and adaptation.

\section*{Acknowledgment}
We thank Hans-Jörg Gisler and Alfonso Blanco for their technical support.
\newpage
\bibliographystyle{IEEEtran} 
\bibliography{references.bib}

\end{document}